\documentclass[prl,aps,twocolumn,showpacs]{revtex4} 
\usepackage{graphicx} 
\usepackage{tabularx}
\usepackage{dcolumn} 
\usepackage{color}
\usepackage{ulem}
\usepackage{bm}
\usepackage{amsmath}
\usepackage{amssymb}

\newcommand{\nn}{\nonumber}

\begin{document} 
 
\title{%
Hidden Crossover Phenomena in Strongly Pauli Limited Multiband Superconductors
---Application to CeCu$_2$Si$_2$---}

\author{Yasumasa Tsutsumi} 
\affiliation{Condensed Matter Theory Laboratory, RIKEN, 
Wako, Saitama 351-0198, Japan} 
\author{Kazushige Machida} 
\affiliation{Department of Physics, Okayama University, 
Okayama 700-8530, Japan} 
\author{Masanori Ichioka} 
\affiliation{Department of Physics, Okayama University, 
Okayama 700-8530, Japan} 

\date{\today}

\begin{abstract}
Motivated by recent experiments on heavy fermion materials
CeCu$_2$Si$_2$ and UBe$_{13}$, we develop a framework to capture generic properties 
of multiband superconductors with strong Pauli paramagnetic effect (PPE).
In contrast to the single band case, the upper critical field $H_{\rm c2}$ can remain second order 
transition even for strong PPE cases. The expected first order transition is hidden inside $H_{\rm c2}$ and becomes a crossover due to the interplay of multibandness. The present theory based on full self-consistent solutions of the microscopic Eilenberger theory explains several mysterious anomalies associated with the crossover
and the ``empty'' vortex core state which is observed by recent STM experiment
on CeCu$_2$Si$_2$.
\end{abstract}

 
 
\maketitle 


There has been much attention focused on multiband superconductors
that were triggered by discoveries of several typical such systems;
MgB$_2$ and iron pnictides in recent years. This leads us to take a fresh look  
not only on new compounds but also on older systems from this multiband perspective.
This new view is particularly fruitful for the oldest heavy fermion superconductors
CeCu$_2$Si$_2$ and UBe$_{13}$, which are key driving materials of this heavy
fermion community over 30 years \cite{sigrist,pfeidler}.
In fact these materials have been regarded as representative
examples of unconventional pairing.
Low $T$ thermodynamics are apparently difficult to understand
within the single band full gap picture and explained in terms of the
nodal gap structure with some power law behaviors. However, it is known
that there is no unique solution for the nodal gap structure so far
because the power law for each thermodynamic quantity  is internally conflicting.
The recent studies on CeCu$_2$Si$_2$ \cite{kittakaCe} and UBe$_{13}$ \cite{shimizu}
raised a strong possibility that they belong to multiband superconductors with
full gaps, which better explains low $T$ thermodynamics than the
unconventional nodal gap model within the single band does.

Pauli paramagnetically limited 
superconductors in the clean limit are characterized by the so-called
Maki parameter $\alpha_{\rm M}\!=\!\sqrt2 H_{\rm c2}^{\rm orb}/H_{\rm P}$, where
$H_{\rm c2}^{\rm orb}$ is the orbital depairing upper critical field and 
$H_{\rm P}\!=\!\Delta_0/\sqrt2 \mu_{\rm B}$ is the Pauli limited field
with the order parameter amplitude $\Delta_0$ at $T\!=\!0$ \cite{fetter,sarma,matsuda}.
If $\alpha_{\rm M}\!\ge\! 1.0$, $H_{\rm c2}$ becomes first order transition (FOT) from second order transition.
Upon further increasing $\alpha_{\rm M}$ above $\alpha_{\rm M}\!\ge\! 1.8$ \cite{GG}, the Fulde-Ferrell-Larkin-Ovchinnikov (FFLO)
state \cite{FF,LO} should appear in the $HT$ phase diagram.
In the strong Pauli limiting $\alpha_{\rm M}\!\rightarrow\!\infty$,
the upper limit temperature of the FOT $T_{\rm 1st}$ along $H_{\rm c2}$
is given by $T_{\rm 1st}/T_{\rm c}\!\rightarrow\! 0.56$.
These important limiting criteria \cite{GG,sarma,matsuda}
often fail badly in some of the heavy fermion superconductors.
For example, CeCu$_2$Si$_2$ and UBe$_{13}$ exhibit a strong rise of $H_{\rm c2}$
at $T_{\rm c}$ ($|dH_{\rm c2}/dT|\!=\!23, 34$ T/K) while $H_{\rm c2}(T\!=\!0)$ is strongly suppressed (2 T, 9 T) respectively.
These numbers give $\alpha_{\rm M}\!=\!3.0$ and 2.3 through a formula for the orbital $H_{\rm c2}$
reduction by Pauli paramagnetic effect (PPE) given in Fig.~1 of Ref.~\cite{machida:2008}, yet the two superconductors show neither FOT nor FFLO phase. 
Thus the  single band picture is fundamentally violated.
Moreover, several outstanding deviations of thermodynamic quantities and local density of states (DOS) from the single band picture are observed.
We summarize the anomalies in the following:

\noindent
(1) The Sommerfeld coefficient $\gamma(H)\!=\!C/T$ in low $T$
shows a kink behavior at $H_{\rm kink}$ above which $\gamma(H)$ starts growing rapidly toward $H_{\rm c2}$ for both compounds~\cite{shimizu,kittakaCe}.
Simultaneously the magnetization curve $M(H)$ in low $T$ gives a minimum~\cite{kittakaCe,shimizu1}.

\noindent
(2) The $T$-dependence of $C/T$ under high fields shows an increasing behavior upon 
lowering $T$ for CeCu$_2$Si$_2$ \cite{kittakaCe}.

\noindent
(3) The empty vortex core state is recently found in CeCu$_2$Si$_2$
by STM \cite{wahl} where the zero energy density of states (ZDOS)
at the core site is suppressed.


The purposes of this paper are to elucidate generic features of the
multiband superconductors with strong Pauli limiting and to place a foundation to
explore these intriguing phenomena. We are going to demonstrate that
FOT is covered by $H_{\rm c2}^{\rm orb}$ in the other band by referring to CeCu$_2$Si$_2$.
We will show that the above items (1)--(3) are generic features of the strong Pauli limited multiband superconductors, reflecting crossover phenomena by the covered FOT and the Zeeman shift of multigap of multiband superconductors.




According to the first principles band calculation in Ref.~\cite{kittakaCe}, the multisheeted Fermi surfaces consist mainly of heavy mass and light mass bands which are physically originated from the hybridization between the atomic $4f$ electrons in Ce atoms and higher $5d$-$6s$ conduction electrons due to Ce and Cu orbitals.
To model them, we consider here a simplified model of a two-band system with a larger superconducting gap band  (band-1) and a smaller gap band (band-2).
For simplicity, the superconducting gap is assumed to open isotropically on each three-dimensional spherical Fermi surface.
Note that the results in this paper are also shown for nodal superconducting gaps~\cite{supplement2}.

The electronic state is calculated by the quasiclassical Eilenberger theory in the clean limit~\cite{eilenberger:1968,ichioka:2004},
including the PPE due to the Zeeman term $\mu B(\bm{r})$~\cite{ichioka}, where $B(\bm{r})$ is the flux density of the internal field and $\mu\!=\!\mu_{\rm B}B_0/\pi k_B T_{\rm c}$ is a renormalized Bohr magneton related to $\alpha_{\rm M}\!=\!1.76\mu$.
The quasiclassical Green's functions $g_j\!\equiv\! g(\bm{k}_j,\bm{r},\omega_n\!+\!i\mu B)$, $f_j\!\equiv\! f(\bm{k}_j,\bm{r},\omega_n\!+\!i\mu B)$, and $\underline{f}_j\!\equiv\!\underline{f}(\bm{k}_j,\bm{r},\omega_n\!+\!i\mu B)$ with band index $j$ depend on the direction of the Fermi momentum $\bm{k}_j$ for each band, the center-of-mass coordinate $\bm{r}$ for the Cooper pair, and Matsubara frequency $\omega_n\!=\!(2n\!+\!1)\pi k_{\rm B}T$ with $n\!\in\!\mathbb{Z}$.
They are calculated in a unit cell of the triangle vortex lattice by solving the Eilenberger equation
\begin{equation}
\begin{split}
&\left\{\omega_n+i\mu B(\bm{r})+\bm{v}_j\cdot\left[\bm{\nabla }+i\bm{A}(\bm{r})\right]\right\}f_j=\Delta_j(\bm{r})g_j,
\\
&\left\{\omega_n+i\mu B(\bm{r})-\bm{v}_j\cdot\left[\bm{\nabla }-i\bm{A}(\bm{r})\right]\right\}\underline{f}_j=\Delta_j^*(\bm{r})g_j,
\end{split}\label{eq:Eilenberger}
\end{equation}
where $g_j\!=\!(1\!-\!f_j\underline{f}_j)^{1/2}$, ${\rm Re}[g_j]\!>\!0$, and $\bm{v}_j\!=\!(v_{{\rm F}j}/v_{{\rm F}0})\bm{k}_j$.
The unit of Fermi velocity $v_{{\rm F}0}$ is defined by $N_{{\rm F}0}v_{{\rm F}0}^2\!\equiv\! N_{{\rm F}1}v_{{\rm F}1}^2\!+\!N_{{\rm F}2}v_{{\rm F}2}^2$, where the DOS in the normal state at each Fermi surface is defined by $N_{{\rm F}0}\!\equiv\! N_{{\rm F}1}\!+\!N_{{\rm F}2}$.
Throughout this paper, temperatures, energies, lengths, and magnetic fields are, respectively, measured in units of the transition temperature $T_{\rm c}$, $\pi k_{\rm B} T_{\rm c}$, $\xi_0\!=\!\hbar v_{{\rm F}0}/2\pi k_{\rm B} T_{\rm c}$, and $B_0\!=\!\phi_0/2\pi\xi_0^2$ ($\phi_0$ is the flux quantum).
%

The gap value is self-consistently calculated by
\begin{align}
\Delta_j(\bm{r})=T\sum_{0<\omega_n\le\omega_{\rm c}}\sum_{j'=1,2}V_{jj'}N_{{\rm F}j'}\left\langle f_{j'}+\underline{f}_{j'}^*\right\rangle_{\bm{k}_{j'}}\label{eq:gap}
\end{align}
where $\langle\cdots\rangle_{\bm{k}_j}$ indicates the Fermi surface average on each band.
We use the energy cutoff $\omega_{\rm c}\!=\!20k_{\rm B}T_{\rm c}$.
The vector potential is also self-consistently determined by
\begin{align}
\bm{\nabla }\!\times\!\bm{\nabla }\!\times\!\bm{A}\!=\!\bm{\nabla }\!\times\!\bm{M}_{\rm para}\!-\!\frac{T}{\tilde{\kappa }^2}\!\sum_{|\omega_n|\le\omega_{\rm c}}\!\sum_{j=1,2}N_{{\rm F}j}\!\left\langle\bm{v}_j{\rm Im}[g_j]\right\rangle_{\bm{k}_j}\!,\label{eq:potential}
\end{align}
which includes the contribution of the paramagnetic moment $\bm{M}_{\rm para}=(0,0,M_{\rm para})$ with
\begin{align}
M_{\rm para}\!=\!M_0\!\left(\!\frac{B(\bm{r})}{\bar{B}}\!-\!\frac{T}{\mu\bar{B}}\!\sum_{|\omega_n|<\omega_{\rm c}}\!\sum_{j=1,2}N_{{\rm F}j}\!\left\langle{\rm Im}[g_j]\right\rangle_{\bm{k}_j}\!\right)\!.\label{eq:Mpara}
\end{align}
The PPE strength is controlled by the Maki parameter $\alpha_{\rm M}\!=\!1.76\mu$.
The normal state paramagnetic moment $M_0\!=\!(\mu/\tilde{\kappa })^2\bar{B}$ and $\tilde{\kappa }\!\equiv\! B_0/(\pi k_{\rm B}T_{\rm c}\sqrt{8\pi N_{{\rm F}0}})\!=\![7\zeta(3)/18]^{1/2}\kappa_{\rm GL}$ with a large Ginzburg-Landau parameter $\kappa_{\rm GL}\!=\!89$.
Using Doria-Gubernatis-Rainer scaling~\cite{doria:1990}, we obtain the relation of the spatial averaged internal field $\bar{B}\!\equiv\!\langle B({\bm r})\rangle_{\bm r}$ and the external field $H$~\cite{ichioka}.
%
%
Then, the total magnetization $M_{\rm total}\!=\!\bar{B}\!-\!H$ including both the diamagnetic and the paramagnetic contributions is derived.

%
%

When we calculate the electronic state, we solve Eq.~\eqref{eq:Eilenberger} with $i\omega_n\!\rightarrow\! E\!+\!i\eta$.
The local density of states (LDOS) is given by $N_j(\bm{r},E)\!=\!N_{j,\uparrow }(\bm{r},E)\!+\!N_{j,\downarrow }(\bm{r},E)$, where
\begin{align}
N_{j,\sigma }(\bm{r},E)\!=\!N_{{\rm F}j}\!\left\langle{\rm Re}\left[g(\bm{k}_j,\bm{r},\omega_n\!+\!i\sigma\mu B)|_{i\omega_n\!\rightarrow\! E\!+\!i\eta }\right]\right\rangle_{\bm{k}_j}\!,
\end{align}
with $\sigma\!=\!1$ $(-1)$ for up (down) spin component.
We typically use the smearing factor $\eta\!=\!0.01$.
The DOS is obtained by the spatial average of the LDOS as $N(E)\!=\!\sum_jN_j(E)\!=\!\sum_j\langle N_{j,\uparrow }(\bm{r},E)\!+\!N_{j,\downarrow }(\bm{r},E)\rangle_{\bm{r}}$.

We set the DOS in the normal state at each Fermi surface to $N_{{\rm F}1}\!=\!\frac{2}{3}N_{{\rm F}0}$ and $N_{{\rm F}2}\!=\!\frac{1}{3}N_{{\rm F}0}$.
We assume that Cooper pair transfer $V_{12}\!=\!V_{21}$ is small.
Then, we set the pairing interaction to $V_{22}=1.5V_{11}$ and $V_{12}=V_{21}=0.05V_{11}$ so that $\Delta_1/\Delta_2\sim 2$ at  zero field.
These two parameters, namely the normal DOS and gap ratios are consistent with the fitting parameters of the specific heat for CeCu$_2$Si$_2$ by the two-gap model~\cite{kittakaCe}.
The band-2 with a small Pauli limited field $H_{\rm P}^{(2)}\!\propto\!\Delta_2$ should have a large orbital limit $H_{\rm c2}^{\rm orb(2)}$ in order that $H_{\rm c2}$ rises sharply at $T_{\rm c}$ as observed in CeCu$_2$Si$_2$;
therefore, we choose the Fermi velocity $v_{{\rm F}1}\!=\!4v_{{\rm F}2}$ giving $H_{\rm c2}^{\rm orb (2)}/H_{\rm c2}^{\rm orb (1)}\!\sim\!4$, where $H_{\rm c2}^{{\rm orb}(j)}\!\propto\!\xi_j^{-2}\!\sim\!\left(\Delta_j/\hbar v_{{\rm F}j}\right)^2$.
When the two bands are independent ($V_{12}\!=\!V_{21}\!=\!0$), the orbital limits $H_{\rm c2}^{{\rm orb}(j)}$ with
$j\!=\!1,2$ are shown schematically by dashed curves in Fig.~1(a),
provided that the transition temperatures are equal which is realized by even slight inter-band interaction.

\begin{figure}
\begin{center}
\includegraphics[width=8cm]{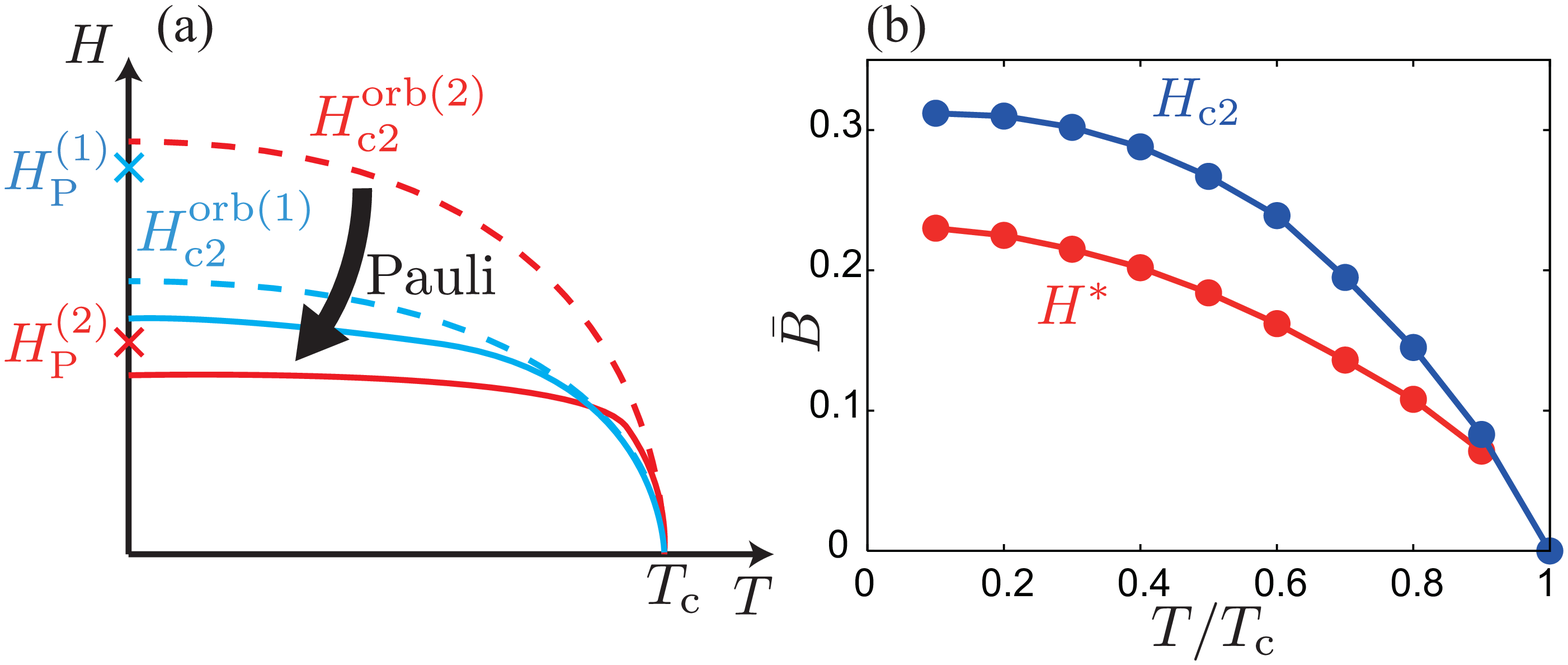}
\end{center}
\caption{\label{fig1}(Color online)
(a) Schematic critical fields. The dashed lines show 
the orbital $H_{\rm c2}^{\rm orb(1)}$ and $H_{\rm c2}^{\rm orb(2)}$.
The solid curves correspond to the
suppressed $H_{\rm c2}^{(1)}$ and $H_{\rm c2}^{(2)}$ by the PPE.
The relative position of the Pauli limited fields $H_{\rm P}^{(1)}$ and $H_{\rm P}^{(2)}$
for each band is shown along the $H$-axis.
(b) Calculated $H_{\rm c2}$ and crossover field $H^{\ast}$
as a function of $T$ for $\alpha_{\rm M}\!=\!1.76$. Note that $H^{\ast}$ terminates at a finite $\bar{B}$,
merging into   $H_{\rm c2}$.
}
\end{figure}


In Fig.~2, we show spatial averaged physical quantities which are probed by thermodynamics.
First, we start off with interacting two band superconductivity without PPE.
As shown in Fig.~2(a-1) the two order parameters $\Delta_1$ and $\Delta_2$
are coupled and vanish at the same $H_{\rm c2}$ when $V_{12}\!=\!V_{21}$ is finite.
Figure 2(a-2) shows the field dependence of the ZDOS, $N(E\!=\!0)$.
At lower fields $N(E\!=\!0)$ grows linearly by $\bar{B}$, which is characteristic to full gap superconductors.
The linear slopes of $N(E\!=\!0)$ for band-1 and band-2 are different reflecting orbital limits $H_{\rm c2}^{\rm orb(1)}\!\approx\!0.5$ and $H_{\rm c2}^{\rm orb(2)}\!\approx\!1.3$~\cite{nakai:2004}.
Note that the ratio of the orbital limits changes from the setting parameter $H_{\rm c2}^{\rm orb (2)}/H_{\rm c2}^{\rm orb (1)}\!\sim\!4$ owing to the inter-band interaction.
The magnetization curve $M(\bar{B})\!=\!M_{\rm total}\!-\!M_0$ is shown in Fig.~2(a-3) which is not much
different from the usual $M(\bar{B})$ curve expected for single band systems.
Those results confirm the naively expected behaviors for two band superconductors.

\begin{figure}
\begin{center}
\includegraphics[width=8.5cm]{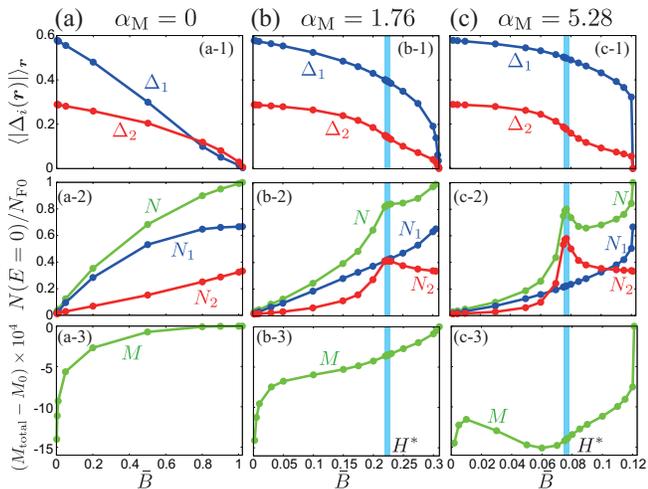}
\end{center}
\caption{\label{fig2}(Color online)
Each column is a series of results for $\alpha_{\rm M}\!=\!0$ (a), $\alpha_{\rm M}\!=\!1.76$ (b), and $\alpha_{\rm M}\!=\!5.28$ (c) at $T\!=\!0.2T_{\rm c}$.
The first row: field dependence of the two order parameters $\Delta_1$ and $\Delta_2$.
The second row: ZDOS $N_1$, $N_2$, and $N$.
The third row: magnetization curve.
Crossover field $H^*$ is shaded in (b) and (c).
Note that the spatial averaged internal field $\bar{B}\equiv\langle B({\bm r})\rangle_{\bm r}$ nearly corresponds to the external field $H$.
}
\end{figure}

Let us now switch on the PPE.
Before going into the numerical results, we
explain an intuitive physical picture.
Since $H_{\rm P}^{(1)}\!>\!H_{\rm P}^{(2)}$ because of the Pauli limited field $H_{\rm P}^{(j)}\!\propto\! \Delta_j$
for the band-$j$, it is expected that the $H_{\rm c2}$ curve for the band-2 is
suppressed much larger than that for the band-1 as schematically shown in Fig.~1(a).
The actual $H_{\rm c2}$ for the two band system is realized by $H_{\rm c2}^{(1)}$
because $H_{\rm c2}^{(2)}$ is less than $H_{\rm c2}^{(1)}$ in low temperatures.
Thus, the $H_{\rm c2}^{(2)}$ curve originally characterized by FOT
is covered by $H_{\rm c2}^{(1)}$.
This expectation is indeed confirmed by our calculation shown in Figs.~2(b) ($\alpha_{\rm M}\!=\!1.76$)
and 2(c) ($\alpha_{\rm M}\!=\!5.28$).

In the $\alpha_{\rm M}\!=\!1.76$ case the resulting $H_{\rm c2}$
is characterized by second order transition. However, inside $H_{\rm c2}$
there exists a crossover field $H^{\ast}$ that corresponds to a kink of $\Delta_2$
upon increasing $\bar{B}$ (Fig.~2(b-1)) and a maximum of ZDOS $N_2(E\!=\!0)$ (Fig.~2(b-2)).
At $H^{\ast}$, $N_2(E\!=\!0)$ exceeds the corresponding normal state value,
that is,  $N_2(E\!=\!0)\!>\!N_{\rm F2}$ whose origin will be explained later.
Then, the total DOS $N(E\!=\!0)$ is also enhanced at $H^{\ast }$ as seen from Fig.~2(b-2).
This feature is indeed observed in CeCu$_2$Si$_2$ 
(see Fig.~2 in Ref.~\cite{kittakaCe}
where $C_e/T$ data at $T\!=\!0.06$ K show an enhancement just below $H_{\rm c2}$)~\cite{supplement}.
Although we assume the $s$-wave pairing state in this numerical calculation, the crossover field $H^*$ with the kink in ZDOS also arises for the nodal $d$-wave pairing state as shown in Supplemental Material~\cite{supplement2}.
The magnetization curve shown in Fig.~2(b-3) exhibits a concave curvature near $H_{\rm c2}$ which is characteristic to the PPE.



Upon further large $\alpha_{\rm M}\!=\!5.28$, $H_{\rm c2}$ becomes eventually FOT from second order transition in low $T$ because $H_{\rm P}^{(1)}\!<\!H_{\rm c2}^{\rm orb(1)}$. As seen from Fig.~2(c-1) in addition
to the crossover field at $H^{\ast}$, also FOT is shown at $H_{\rm c2}\!=\!0.12$ where the two order parameters vanish suddenly.
In low $\bar{B}$ region the ZDOS is strongly suppressed and exhibits successively
sharp variations at $H^{\ast}$ and $H_{\rm c2}$ as seen from Fig.~2(c-2).
The magnetization curve in Fig.~2(c-3) exhibits a minimum just below $H^{\ast}$,
which is caused by the competition between the orbital diamagnetic negative contribution
and the paramagnetic positive contribution due to the PPE.
This minimum is observed both in CeCu$_2$Si$_2$
(see Fig.~4(b) in Ref.~\cite{kittakaCe}) and UBe$_{13}$ (see Fig.~2 in Ref.~\cite{shimizu1}).
Note that the minimum of the magnetization curve never occurs in the single band case~\cite{ichioka}.

The $HT$ phase diagram is shown in Fig.~1(b) where
the second order transition $H_{\rm c2}$ and the crossover field $H^{\ast}$
are depicted for the $\alpha_{\rm M}\!=\!1.76$ case.
It is seen from Fig.~1(b) that $H^{\ast}$ terminates at a finite $\bar{B}$ hit on the $H_{\rm c2}$ curve
because the PPE becomes effective at a finite $\bar{B}$.
We note that the crossover field $H^{\ast}$ relative to $H_{\rm c2}$ shown in Fig.~1(b) looks very similar to $H_{\rm Mag}^{\ast }$ observed in UBe$_{13}$~\cite{shimizu1}, although the BCS theory with a constant DOS at the Fermi level is a simple approximation for UBe$_{13}$, which shows non-Fermi-liquid behaviors.

\begin{figure}
\begin{center}
\includegraphics[width=7cm]{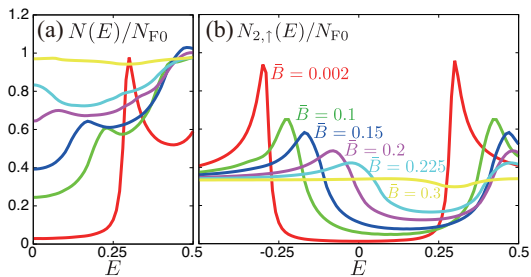}
\end{center}
\caption{\label{fig3}(Color online)
Total DOS $N(E)$ (a) and DOS for up spin component in band-2 $N_{2,\uparrow }(E)$ (b) for various $\bar{B}$ with $\alpha_{\rm M}\!=\!1.76$ at $T\!=\!0.2T_{\rm c}$.
At $\bar{B}\!=\!0.002$, $\Delta_1\!=\!0.58$ and $\Delta_2\!=\!0.29$.
By increasing $\bar{B}$, the gap edge singularities corresponding to the minor gap 
$\Delta_2$ move inward to $E\!=\!0$. At $\bar{B}\!=\!0.225$, total DOS has a maximum at $E\!=\!0$.
}
\end{figure}

The spatial averaged DOS $N(E)$ calculated 
for various $\bar{B}$ in the $\alpha_{\rm M}\!=\!1.76$ case  
is shown in Fig.~3(a).
At $\bar{B}\!=\!0.002$ the gap due to $\Delta_1\!=\!0.58$ and $\Delta_2\!=\!0.29$ widely opens and 
$N(E)$ exhibits sharp edge singularities at $\Delta_2$.
Upon increasing $\bar{B}$ from $\bar{B}\!=\!0.002$ toward $H_{\rm c2}\!\approx\! 0.3$
it is seen that ZDOS increases gradually due to the depaired quasiparticles
in the vortex core region~\cite{ichioka}.
Finally at $\bar{B}\!=\!0.225$ below $H_{\rm c2}$ corresponding to the crossover field
$H^{\ast}$, $N(E)$ exhibits a maximum at $E\!=\!0$
because the gap edge singularity of the minor band gap $\Delta_2$ is shifted to $E\!=\!0$ by the PPE.
The Zeeman shifted DOS for up spin component in band-2 is shown in Fig.~3(b) whose energy is shifted by $\mu B$ ($\alpha_{\rm M}\!=\!1.76$ corresponds to $\mu\!=\!1$).
Note that the DOS for the down spin component has a relation $N_{j,\downarrow }(E)\!=\!N_{j,\uparrow }(-E)$.
The Zeeman shifted gap edge singularity at $E\!=\!\Delta_2\!-\!\mu H^{\ast }\!=\!0$ explains also the enhanced $N_2(E\!=\!0)$ behaviors at $H^{\ast}$ shown in Fig.~2(b-2).

Low energy DOS can be observed as $C(T)/T$ in low temperatures with an appropriate scale transformation from $k_{\rm B}T$ to $E$ as demonstrated in Supplemental Material \cite{supplement}.
In CeCu$_2$Si$_2$, the observed electronic specific heat $C_{\rm e}/T$ increases toward low temperature under high fields (Fig.~1 in Ref.~\cite{kittakaCe}), which will reflect the enhancement of the DOS toward $E\!=\!0$ at $H^{\ast }$.

So far we have mainly discussed the averaged physical quantities.
Now we touch upon the local ones that are also important 
to characterize the crossover phenomena associated with the multiband Pauli limited 
superconductors.
In Fig.~4 we show the local ZDOS around the vortex core, each corresponding
to major band-1 (Fig.~4(a)) and minor band-2 (Fig.~4(b)).
The ring shaped ``crater" like landscapes are clearly seen, which are quite different 
from that of the ordinary cases without the PPE where the single peaked mountain like landscape
is seen~\cite{hayashi1,hayashi2}.

\begin{figure}
\begin{center}
\includegraphics[width=8.5cm]{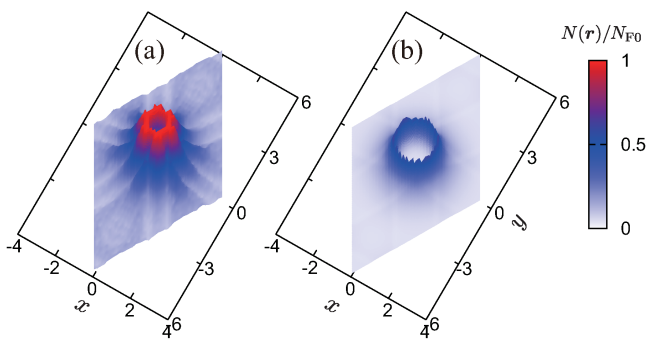}
\end{center}
\caption{\label{fig4}(Color online)
Local ZDOS landscapes around a vortex core for the band-1 (a) and 
band-2 (b) with $\mu\!=\!1$ at $\bar{B}\!=\!0.15$ and $T\!=\!0.2T_{\rm c}$.
One unit cell of the triangular vortex lattice is shown with the length measured by $\xi_0$. 
}
\end{figure}

The physical origin of this structure of ZDOS can be understood by the Zeeman shift of the vortex bound state as follows.
At the ordinary vortex core site the vortex bound state has a peak 
exactly at $E\!=\!0$.
As moving away from the core this peak splits into two peaks 
(see Fig.~9 in Ref.~\cite{hayashi1}), which are eventually absorbed and merged into the continuum above the gap edges.
In the cases with the PPE, the Zeeman split two peaks at the core site evolve into two peaks each when moving away.
These inward two peaks intersect somewhere away from the core. Thus these particular sites situated circularly
give rise to the peak, resulting in a ring structure in the ZDOS landscapes. This peak position $r_{\rm max}$ is roughly estimated by $r_{\rm max}/\xi_0\!\sim\! E_B/\Delta_0$ with the Zeeman shift $E_B$ from $E\!=\!0$.

The empty core with the crater like landscape is actually observed by a recent STM
experiment on CeCu$_2$Si$_2$~\cite{wahl}.
Observed $r_{\rm max}/\xi_0\!\sim\!0.5$ by the experiment at $H\!=\!1.6$ T implies
$E_B/\Delta_0\!\sim\!0.5$. The Zeeman energy can be directly checked by future STM 
experiment at the core site where we expect that the split two peaks 
are observed at the half energy of the gap.


Although we have focused on heavy fermion superconductors CeCu$_2$Si$_2$ and UBe$_{13}$ in this paper,
our conclusions also apply to other strongly Pauli limited multiband superconductors, e.g.~KFe$_2$As$_2$ and FeSe belonging to iron pnictides family~\cite{lei}.
An ordinary FOT was observed in KFe$_2$As$_2$~\cite{zocco}, however, which also shows a kink of $\gamma(H)\!=\!C/T$ in low $T$ and an increase of $C(T)/T$ upon lowering $T$ under high fields~\cite{kittakaFe} similarly to CeCu$_2$Si$_2$.
For FeSe without the FOT, the thermal conductivity anomaly was observed under high fields~\cite{kasahara:arXiv}.
Since thermal conductivity depends on the quasiparticle structure, the anomaly may result from the enhancement of ZDOS at the crossover field.

In summary, we have constructed a general framework to describe the Pauli paramagnetic effect (PPE)
for multiband superconductors within microscopic Eilenberger theory applicable to most type II superconductors.
The present theory yields a better and advanced understanding for a superconductor with PPE than those firmly established
frameworks~\cite{fetter,sarma,GG} based on the single band assumption.
We applied it to CeCu$_2$Si$_2$ and interpreted generic features (1)--(3) for the strong Pauli limited multiband superconductors.
We showed the crossover phenomena at $H^{\ast}$ deep inside $H_{\rm c2}$ in the $HT$ plane
with (1) an enhancement of the Sommerfeld coefficient $\gamma(H)$ and a minimum of the magnetization curve.
This feature is also observed in UBe$_{13}$.
(2) An increase of low temperature specific heat and (3) empty core vortices observed in CeCu$_2$Si$_2$ are due to the Zeeman shift of multigap.
The crossover phenomena are generic features of the strong Pauli limited multiband superconductors irrespective of the gap structure~\cite{supplement2}.
However, multiband superconductor CeCoIn$_5$ shows a usual FOT at $H_{\rm c2}$ or the FFLO state under high fields~\cite{matsuda} because CeCoIn$_5$ is effectively regarded as a single band superconductor with strong PPE.
The smaller gap band in CeCoIn$_5$ is readily reached to the orbital limit owing to $H_{\rm c2}^{\rm orb(2)}\ll H_{\rm c2}^{\rm orb(1)}$~\cite{seyfarth:2008,allan:2013}.
Our multiband picture will give a hint why the FFLO is difficult to realize in large $\alpha_{\rm M}$ superconductors.
Note that the FFLO phase qualitatively changes for multiband superconductors~\cite{mizushima:2014, gurevich:2011} even without the crossover phenomena.



We thank T.~Sakakibara, S.~Kittaka, Y.~Shimizu, P.~Wahl and N.~Nakai for helpful discussions.
This work is supported by Grant-in-Aid for Scientific Research No.~26400360 and No.~25103716 from the Japan Society for the Promotion of Science.


\clearpage
\onecolumngrid

\renewcommand{\thefigure}{S\arabic{figure}} 

\renewcommand{\thetable}{S\arabic{table}} 

\renewcommand{\thesection}{S\arabic{section}.}

\renewcommand{\theequation}{S.\arabic{equation}}

\setcounter{figure}{0}
\setcounter{table}{0}
\setcounter{equation}{0}

\begin{flushleft} 
{\Large {\bf Supplementary Material}}
\end{flushleft}

\baselineskip24pt

\begin{flushleft} 
{\bf S1. Crossover phenomena in $d$-wave pairing state}
\end{flushleft} 

\begin{figure}
\begin{center}
\includegraphics[width=10cm]{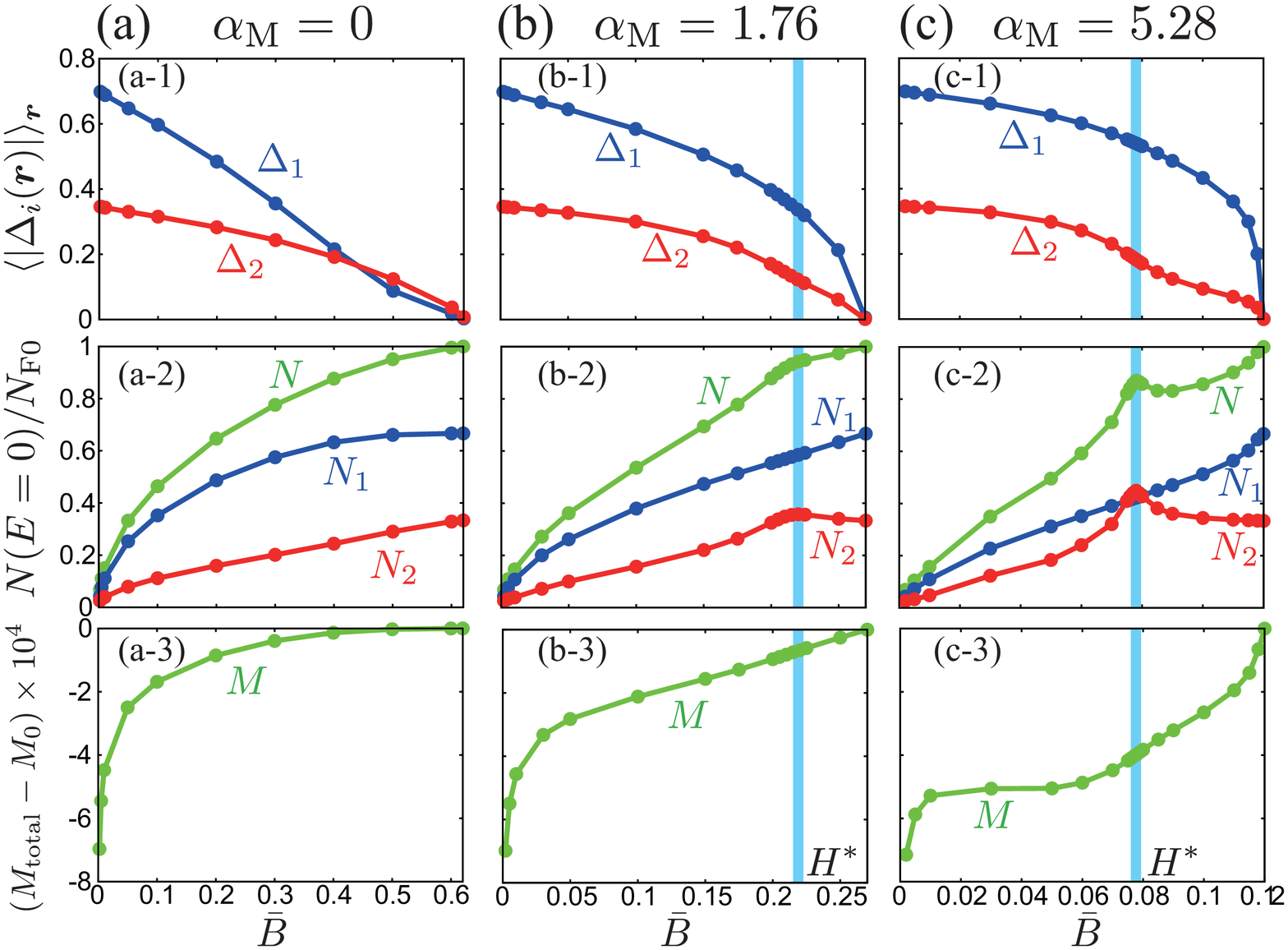}
\end{center}
\caption{\label{fig1s}(Color online)
Each column is a series of results for $\alpha_{\rm M}\!=\!0$ (a), $\alpha_{\rm M}\!=\!1.76$ (b), and $\alpha_{\rm M}\!=\!5.28$ (c) at $T\!=\!0.2T_{\rm c}$ for the $d$-wave pairing state.
The first row: spatial averaged internal field dependence of the two order parameters $\Delta_1$ and $\Delta_2$.
The second row: ZDOS $N_1$, $N_2$, and $N$.
The third row: magnetization curve.
Crossover field $H^*$ is shaded in (b) and (c).
}
\end{figure}

We show spatial averaged order parameter, zero energy density of states (ZDOS), and magnetization for the $d$-wave pairing state in Fig.~\ref{fig1s}.
We consider $d_{x^2-y^2}$-wave gap function, $\Delta_j({\bm k}_j,{\bm r})\equiv\Delta_j({\bm r})\phi({\bm k}_j)$, with $\phi({\bm k})=\sqrt{2}(k_x^2-k_y^2)$ on the two-band cylindrical Fermi surface.
Quasiclassical Green's functions for anisotropic pairing states are self-consistently calculated by the Eilenberger equation~\cite{machida:2008b}
\begin{equation}
\begin{split}
&\left\{\omega_n+i\mu B(\bm{r})+\bm{v}_j\cdot\left[\bm{\nabla }+i\bm{A}(\bm{r})\right]\right\}f_j=\Delta_j(\bm{r})\phi({\bm k}_j)g_j,
\\
&\left\{\omega_n+i\mu B(\bm{r})-\bm{v}_j\cdot\left[\bm{\nabla }-i\bm{A}(\bm{r})\right]\right\}\underline{f}_j=\Delta_j^*(\bm{r})\phi({\bm k}_j)g_j,
\end{split}\label{eqs:Eilenberger}
\end{equation}
with the gap equation
\begin{align}
\Delta_j(\bm{r})=T\sum_{0<\omega_n\le\omega_{\rm c}}\sum_{j'=1,2}V_{jj'}N_{{\rm F}j'}\left\langle \phi^*({\bm k}_{j'})\left(f_{j'}+\underline{f}_{j'}^*\right)\right\rangle_{\bm{k}_{j'}}.
\label{eqs:gap}
\end{align}
For the calculation in Fig.~\ref{fig1s}, we use the same parameters for the calculation in Fig.~2 of the main text, i.e., the density of states in the normal state $2N_{\rm F1}=N_{\rm F2}=\frac{1}{3}N_{\rm F0}$, the Fermi velocity $v_{\rm F1}=4v_{\rm F2}$, and the pairing interaction $V_{22}=1.5V_{11}$ and $V_{12}=V_{21}=0.05V_{11}$.

According to our setting $H_{\rm c2}^{\rm orb(2)}>H_{\rm c2}^{\rm orb(1)}$, the order parameter $\Delta_1$ is strongly suppressed by the magnetic field without Pauli paramagnetic effect (PPE) as shown in Fig.~\ref{fig1s}(a-1).
Figure \ref{fig1s}(a-2) shows the field dependence of the ZDOS, $N(E=0)$, which grows by $\sqrt{\bar{B}}$ in low fields due to the nodal excitations~\cite{volovik:1993b,ichioka:1999b}.
By the rapid rise of the low-energy excitations, the upper critical field of the $d$-wave pairing state at $H_{\rm c2}\approx0.6$ is lower than that of the $s$-wave pairing state at $H_{\rm c2}\approx1.0$ as shown in Fig. 2(a) of the main text.

When we turn on the PPE, $H_{\rm c2}$ is suppressed to $H_{\rm c2}\approx0.3$ for $\alpha_{\rm M}=1.76$ in Fig.~\ref{fig1s}(b) and to $H_{\rm c2}\approx0.12$ for $\alpha_{\rm M}=5.28$ in Fig.~\ref{fig1s}(c).
The upper critical fields are strongly subjected to the PPE because they nearly correspond with these in the $s$-wave pairing state in spite of the difference of $H_{\rm c2}^{\rm orb}$.
Below $H_{\rm c2}$ there exist a crossover field $H^*$ where $N_2(E=0)$ exceeds the normal state value [Figs.~\ref{fig1s}(b-2) and \ref{fig1s}(c-2)].
The crossover fields nearly correspond with these in the $s$-wave pairing state, which implies that not the gap structure but the multiband nature is important to the crossover phenomenon.
The influence of the nodal gap structure only weakens the enhancement of the ZDOS at $H^*$ by low-energy excitations from nodes.
Since a plateau in the magnetization curve $M(\bar{B})=M_{\rm total}-M_0$ has been seen in Fig.~\ref{fig1s}(c-3), a local minimum in the magnetization curve will arise in larger Maki parameter cases, $\alpha_{\rm M}>5.28$.

\begin{flushleft} 
{\bf S2. Transformation from $C(T)/T$ to $N(E)$}
\end{flushleft} 

Specific heat is given by a temperature derivative of entropy as
\begin{align}
C=&T\frac{dS}{dT}
=\sum_{\bm{k}}E_{\bm{k}}\frac{\partial f_{\bm{k}}}{\partial T}\nn\\
=&\int_{-\infty }^{\infty }E\frac{k_{\rm B}\beta^2Ee^{\beta E}}{(e^{\beta E}+1)^2}N(E)dE,
\end{align}
where $f_{\bm{k}}=(e^{\beta E_{\bm k}}+1)^{-1}$ is the Fermi distribution function with $\beta=1/k_{\rm B}T$.
We transform a variable of the integral from $E$ to $k_{\rm B}Tx$; then, the specific heat is described by
\begin{align}
\frac{C}{T}=k_{\rm B}^2\int_{-\infty }^{\infty }\frac{x^2e^x}{(e^x+1)^2}N(k_{\rm B}Tx)dx.
\end{align}
If low energy DOS can be expanded to $N(E)=N(E=0)+A|E|^{\alpha }$ for $\alpha>0$, the specific heat in low temperature is expanded to
\begin{align}
\frac{C}{T}=&k_{\rm B}^2\int_{-\infty }^{\infty }\frac{x^2e^x}{(e^x+1)^2}[N(0)+A|k_{\rm B}Tx|^{\alpha }]dx\nn\\
=&k_{\rm B}^2N(0)\int_{-\infty }^{\infty }\frac{x^2e^x}{(e^x+1)^2}dx+k_{\rm B}^2A|k_{\rm B}T|^{\alpha }\int_{-\infty }^{\infty }\frac{|x|^{\alpha }x^2e^x}{(e^x+1)^2}dx\nn\\
=&B_0k_{\rm B}^2N(0)+B_{\alpha }k_{\rm B}^2A|k_{\rm B}T|^{\alpha }\nn\\
=&B_0k_{\rm B}^2\left[N(0)+\frac{B_{\alpha }}{B_0}A|k_{\rm B}T|^{\alpha }\right]\nn\\
=&B_0k_{\rm B}^2N(k_{\rm B}T'),
\end{align}
where $T'=\left(\frac{B_{\alpha }}{B_0}\right)^{1/\alpha }T$ and
\begin{align}
B_{\alpha }=&2\int_0^{\infty }\frac{x^{\alpha+2}e^x}{(e^x+1)^2}dx\nn\\
=&(2-2^{-\alpha })(\alpha+2)\Gamma(\alpha+2)\zeta(\alpha+2),\nn
\end{align}
particularly, $B_0=\pi^2/3$, $B_1=9\zeta(3)$, and $B_2=7\pi^4/15$.
Therefore, $C/T$ in low temperature gives low energy DOS after a scale transformation from $k_{\rm B}T$ to $E$.

\end{document}